\documentclass[fleqn,usenatbib]{mnras}

\usepackage{newtxtext,newtxmath}
 
\usepackage[T1]{fontenc}

\DeclareRobustCommand{\VAN}[3]{#2}
\let\VANthebibliography\thebibliography
\def\thebibliography{\DeclareRobustCommand{\VAN}[3]{##3}\VANthebibliography}

\usepackage{graphicx}	
\usepackage{amsmath}	
\usepackage{amssymb}	
\usepackage{multirow}

\title[High-resolution TRAPPIST-1e spectral models]{High-Resolution spectral models of TRAPPIST-1e seen as a \textit{Pale Blue Dot} for ELT and \textit{JWST} observations}

\author[Lin and Kaltenegger]{
Zifan Lin$^{1, 2}$\thanks{E-mail: zifanlin@mit.edu} and
Lisa Kaltenegger$^{2}$
\\
$^{1}$Department of Earth, Atmospheric and Planetary Sciences, Massachusetts Institute of Technology, Cambridge, MA 02139, USA\\
$^{2}$Carl Sagan Institute and Department of Astronomy, Cornell University, Ithaca, NY 14853, USA
}

\date{Accepted XXX. Received YYY; in original form ZZZ}

\pubyear{2021}

\begin{document}
\label{firstpage}
\pagerange{\pageref{firstpage}--\pageref{lastpage}}
\maketitle

\begin{abstract}
Rocky exoplanets orbiting in the habitable zone (HZ) of nearby M dwarfs provide unique opportunities for characterizing their atmospheres and searching for biosignature gases. TRAPPIST-1e, a temperate Earth-sized exoplanet in the HZ of a nearby M dwarf, is arguably the most favorable target for ground- and space-based atmospheric characterization by the extremely large telescopes (ELTs) and the \textit{James Webb Space Telescope} (\textit{JWST}). To inform future observations in reflected and emitted lights using these upcoming telescopes, we simulate the high-resolution reflection and emission spectra for TRAPPIST-1e for both modern and prebiotic Earth-like atmospheric compositions. To demonstrate the effects of wavelength-dependent albedo on climate and spectra, we further simulate five albedo scenarios for each atmospheric composition: cloudy modern Earth-like, cloud-free modern Earth-like, cloudy ocean planet, 100 per cent cloudy planet, and wavelength-independent albedo of 0.31. We use the recent Mega-MUSCLES spectral energy distribution (SED) of TRAPPIST-1 for our models. We show that the O$_2$ + CH$_4$ and O$_3$ + CH$_4$ biosignature pairs as well as climate indicators (CO$_2$ and H$_2$O) show features in both high-resolution reflection and emission spectra of TRAPPIST-1e that the ELTs can search for. Our high-resolution database for modern and prebiotic Earth TRAPPIST-1e models with various surface compositions and cloud distributions provides a tool for observers to train retrieval algorithms and plan observation strategies to characterize this potentially habitable world.
\end{abstract}




\section{Introduction} \label{sec:intro}

To date, over 5000 exoplanets in more than 3500 planetary systems have been detected.\footnote{\href{exoplanets.nasa.gov}{exoplanets.nasa.gov}} Some of these planets reside in the circumstellar HZ \citep[see e.g.,][]{Quintana_2014Sci...344..277Q, Kane_2016ApJ...830....1K, kaltenegger_how_2017, Johns_2018ApJS..239...14J, Berger_2020AJ....159..280B}, a region in which the radiation from one or multiple host stars allows for liquid water on the surface of terrestrial planets \citep[see e.g.,][]{Kasting_1993Icar..101..108K}. Direct imaging allows remote characterization of such potentially habitable worlds and the search for biosignature gases on them: to view them as \textit{Pale Blue Dots}. Their reflected and emitted light contain the information on their atmospheric and surface compositions \citep[e.g.,][]{Sagan_1993Natur.365..715S, kaltenegger_how_2017, Schwieterman_2018AsBio..18..663S}.

Due to the small stellar radius and close-in HZ of M dwarfs, the planet-to-star contrast ratio ($F_{\rm p}/F_*$) for both reflection and emission spectra of a M dwarf exoplanet are greater than other stellar types, assuming the same planetary radius. In addition, occurrence frequency of Earth-size planets orbiting M dwarfs appear to be high \citep[see e.g.,][]{scalo_m_2007, dressing_occurrence_2015}, and M dwarfs make up about 75 per cent of all stars in proximity to the Sun. 

Several Earth-sized planets in the HZ of M dwarfs have been discovered in the solar neighbourhood \citep[e.g.,][]{anglada-escude_terrestrial_2016, dittmann_temperate_2017, bonfils_temperate_2018}. Among these systems, the TRAPPIST-1 system at 12.1 pc hosts seven transiting Earth-sized exoplanets, with four planets (TRAPPIST-1d, e, f, and g) in or near the HZ \citep{gillon_temperate_2016, gillon_seven_2017}. The TRAPPIST-1 system provides a unique opportunity for studying seven temperate rocky exoplanets that formed and evolved in the same planetary system \citep{Gillon_2020arXiv200204798G}. TRAPPIST-1 is an intriguing target system for upcoming ground-based ELTs and space-based telescopes such as the \textit{JWST}. We focus on TRAPPIST-1e here because of its prime location in the middle of the HZ, which allows it to maintain a surface temperature above freezing with a slight increase in greenhouse gas -- here we use 100$\times$ present atmospheric level (PAL) CO$_2$ concentration (see \citealt{Lin_2021MNRAS.505.3562L} for model details).
The surface habitability and detectability of atmospheric spectral features of TRAPPIST-1e has been assessed from multiple perspectives, utilizing models ranging from Venus-like acid atmospheres to Earth-like N$_2$-dominated atmosphere containing O$_2$ and CO$_2$ \citep[e.g.,][]{wolf_assessing_2017, dong_atmospheric_2018, lincowski_evolved_2018, turbet_modeling_2018, fauchez_impact_2019, Lustig-Yaeger2019, omalley-james_lessons_2019, Lin_2021MNRAS.505.3562L}, as well as model that couples atmosphere and interior evolution over Gyr timescale \citep{krissansen-totton_predictions_2022}.

\begin{figure*}
	\includegraphics[width=\textwidth]{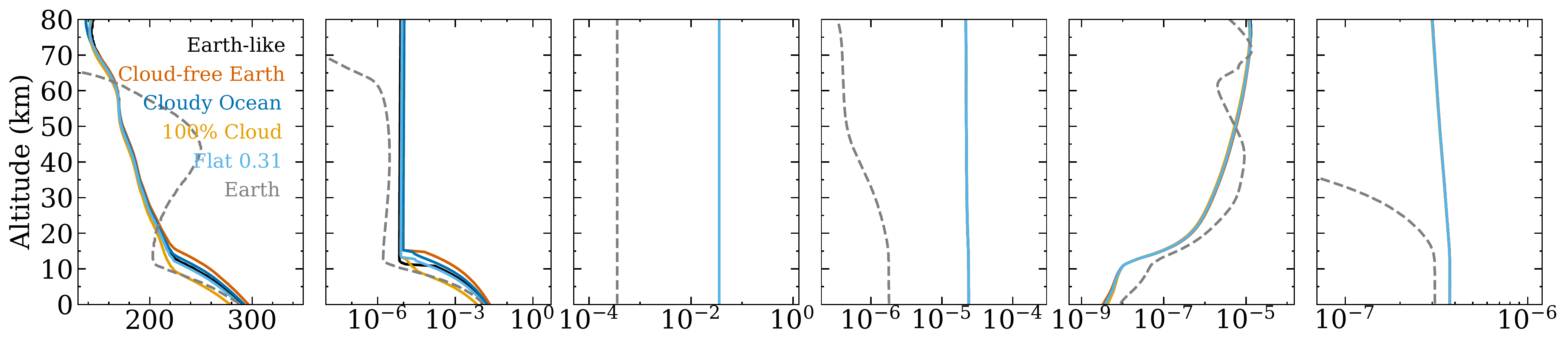}
	\includegraphics[width=\textwidth]{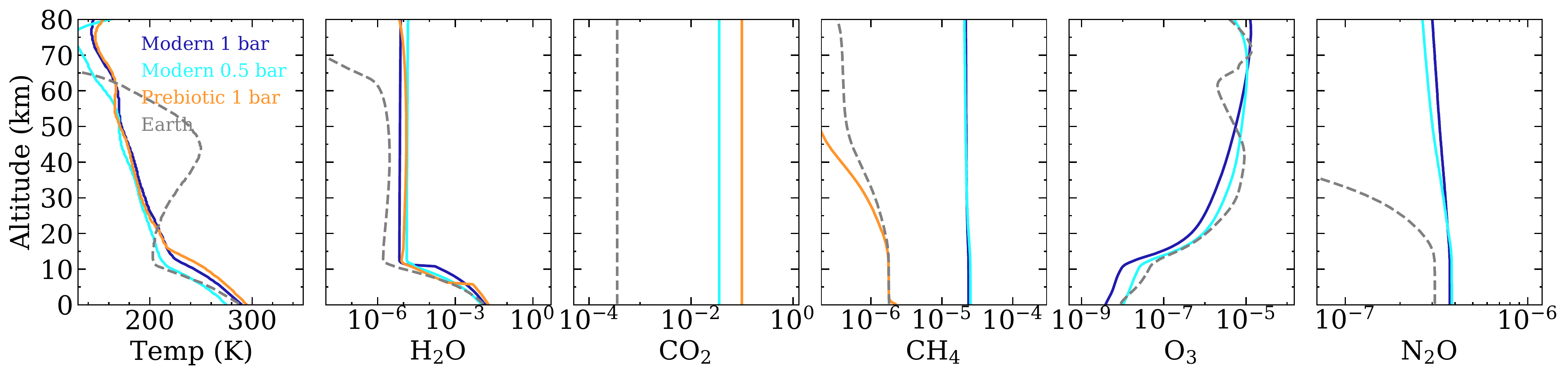}
    \caption{Temperature profiles and mixing ratio profiles of major chemicals in our TRAPPIST-1e model atmospheres. The top row show modern Earth-like models with 1 bar surface pressure with four surface compositions: (black) modern Earth-like surface with 44 per cent cloud coverage, (red) cloud-free modern Earth-like surface, (blue) ocean planet with 44 per cent cloud coverage, (yellow) a planet fully covered by clouds, and (sky blue) a planet with wavelength-independent albedo of 0.31. The bottom row shows three model Earth-like surface models with different surface pressures and atmospheric compositions: (navy) modern Earth-like atmosphere with 1 bar surface pressure, (cyan) eroded modern Earth-like atmosphere with 0.5 bar surface pressure, and (orange) prebiotic Earth-like composition. Profiles of modern Earth are overplotted for comparison (gray dashed lines). In the prebiotic atmosphere, O$_3$ and N$_2$O have negligible mixing ratio below the plotting limits.}
    \label{fig:temp_mr_profile}
\end{figure*}

A previous paper by the authors \citep{lin_high-resolution_2019} demonstrated that the reflection spectra of TRAPPIST-1e show features of important climate indicators (CO$_2$ and H$_2$O) as well as O$_2$ + CH$_4$ and O$_3$ + CH$_4$, which are considered as strong biosignature pairs \citep[see e.g.,][]{Lederberg_1965Natur.207....9L, Lovelock_1965Natur.207..568L, Lippincott_1967ApJ...147..753L, Sagan_1993Natur.365..715S}. However, since the publication of the previous paper, improvements have been made both to our models and to the host star spectrum, requiring an update to provide the best tool for observers to train retrieval algorithms and plan observing strategies. 

First, a semiempirical SED of TRAPPIST-1 from the Mega-MUSCLES (Measurements of the Ultraviolet Spectral Characteristics of Low-Mass Exoplanetary Systems) \textit{Hubble Space Telescope} (\textit{HST}) treasury program was recently published \citep{Wilson_2021ApJ...911...18W}. \textsc{Exo-Prime2}, our one-dimensional climate, photochemistry, and spectral simulation model, has also been updated. The model now fully incorporates the effects of wavelength-dependent surface albedo (described in detail in \citealt{Madden_2020_surfaces}). As shown in \cite{lin_high-resolution_2019}, different surface composition and cloud coverage can greatly change the intensity of reflected light as a function of wavelength. Consequently, the wavelength-dependent surface albedo of the planet can influence the energy balance and hence climate of a planet as well as its spectrum. 

In this paper, we present updated high-resolution reflection spectra from 0.4 to 5 $\micron$ and emission spectra from 4 to 20 $\micron$ for TRAPPIST-1e based on the improved models. All spectra are available online\footnote{\href{https://doi.org/10.5281/zenodo.6365743}{https://doi.org/10.5281/zenodo.6365743}} to provide forward model inputs for observational and retrieval simulations of TRAPPIST-1e with telescopes like \textit{JWST} and E-ELT, as well as future concepts like HabEx \citep{habex_2016SPIE.9904E..0LM}, LUVOIR \citep{LUVOIR_2018arXiv180909668T}, and Origins \citep{Origins_2018NatAs...2..596B}.


\begin{table*}
\caption{Model parameters at the surface. Both modern Earth-like models have 100 times more CO$_2$ compared to PAL to keep the coolest, 0.5 bar model above freezing at the surface when assuming an Earth-like albedo.}
\label{tab:surface_parameters}
\begin{tabular}{llllllllll}
\hline
\multirow{2}{*}{Epoch} & \multirow{2}{*}{Albedo} & \multirow{2}{*}{P$_{\textrm{surf}}$ (bar)} & \multirow{2}{*}{T$_{\textrm{surf}}$ (K)} & \multicolumn{6}{c}{Surface mixing ratio} \\
   &  &  &   & CO$_2$ & H$_2$O  & O$_2$ & O$_3$  & CH$_4$  & N$_2$O \\ 
\hline
\multirow{5}{*}{Modern} & Earth-like & 1.0 & 289   & $3.60 \times 10^{-2}$ & $1.36 \times 10^{-2}$ & 0.21 & $3.90 \times 10^{-9}$  & $2.38 \times 10^{-5}$ & $3.74 \times 10^{-7}$  \\
& Cloud-free Earth & 1.0 & 296   & $3.60 \times 10^{-2}$ & $2.06 \times 10^{-2}$ & 0.21 & $3.41 \times 10^{-9}$  & $2.39 \times 10^{-5}$ & $3.75 \times 10^{-7}$  \\
& Cloudy Ocean & 1.0 & 292   & $3.60 \times 10^{-2}$ & $1.61 \times 10^{-2}$ & 0.21 & $3.80 \times 10^{-9}$  & $2.39 \times 10^{-5}$ & $3.74 \times 10^{-7}$  \\
& 100\% Cloud & 1.0 & 278   & $3.60 \times 10^{-2}$ & $6.40 \times 10^{-3}$ & 0.21 & $4.41 \times 10^{-9}$  & $2.38 \times 10^{-5}$ & $3.74 \times 10^{-7}$  \\
& Flat 0.31 & 1.0 & 287   & $3.60 \times 10^{-2}$ & $1.22 \times 10^{-2}$ & 0.21 & $3.86 \times 10^{-9}$  & $2.38 \times 10^{-5}$ & $3.74 \times 10^{-7}$  \\ 
\hline
\multirow{5}{*}{Modern} & Earth-like & 0.5 & 274   & $3.60 \times 10^{-2}$ & $1.00 \times 10^{-2}$ & 0.21 & $1.05 \times 10^{-8}$  & $2.53 \times 10^{-5}$ & $3.84 \times 10^{-7}$ \\
& Cloud-free Earth & 0.5 & 282   & $3.60 \times 10^{-2}$ & $1.78 \times 10^{-2}$ & 0.21 & $9.44 \times 10^{-9}$  & $2.58 \times 10^{-5}$ & $3.86 \times 10^{-7}$  \\
& Cloudy Ocean & 0.5 & 277   & $3.60 \times 10^{-2}$ & $1.26 \times 10^{-2}$ & 0.21 & $1.04 \times 10^{-8}$  & $2.55 \times 10^{-5}$ & $3.84 \times 10^{-7}$  \\
& 100\% Cloud & 0.5 & 259   & $3.60 \times 10^{-2}$ & $2.66 \times 10^{-3}$ & 0.21 & $1.14 \times 10^{-8}$  & $2.53 \times 10^{-5}$ & $3.84 \times 10^{-7}$  \\
& Flat 0.31 & 0.5 & 271   & $3.60 \times 10^{-2}$ & $7.58 \times 10^{-3}$ & 0.21 & $1.04 \times 10^{-8}$  & $2.53 \times 10^{-5}$ & $3.84 \times 10^{-7}$ \\ 
\hline
\multirow{5}{*}{Prebiotic} & Earth-like  & 1.0 & 294   & 0.1 & $1.84 \times 10^{-2}$ & $10^{-13}$ & $2.75 \times 10^{-17}$ & $2.20 \times 10^{-6}$ & $5.68 \times 10^{-11}$ \\
& Cloud-free Earth & -- & --   & -- & -- & --  & -- & -- & --  \\
& Cloudy Ocean & 1.0 & 289   & 0.1 & $1.39 \times 10^{-2}$ & $10^{-13}$ & $2.82 \times 10^{-17}$  & $2.20 \times 10^{-6}$ & $5.68 \times 10^{-11}$  \\
& 100\% Cloud & 1.0 & 288   & 0.1 & $1.26 \times 10^{-2}$ & $10^{-13}$ & $1.94 \times 10^{-17}$  & $2.20 \times 10^{-6}$ & $5.68 \times 10^{-11}$  \\
& Flat 0.31 & 1.0 & 288   & 0.1 & $1.27 \times 10^{-2}$ & $10^{-13}$ & $2.77 \times 10^{-17}$  & $2.20 \times 10^{-6}$ & $5.68 \times 10^{-11}$ \\ 
\hline
\end{tabular}
\end{table*}

\section{Methods} \label{sec:methods}
Our TRAPPIST-1 stellar model is based on observations of the Mega-MUSCLES program \citep{Wilson_2021ApJ...911...18W}, an extension to the original MUSCLES \textit{HST} Treasury program \citep{france_muscles_2016, Youngblood_2016ApJ...824..101Y, loyd_muscles_2018}. We scale the stellar spectrum to the irradiation at the top of atmosphere of TRAPPIST-1e by the squared ratio of TRAPPIST-1’s distance from Earth (12.425 pc, following \textit{Gaia} DR2 parallax presented by \citealt{Bailer-Jones_2018AJ....156...58B}) to TRAPPIST-1e’s distance from its host (0.02925 au, following \citealt{agol_refining_2021}).

We use \textsc{Exo-Prime2}, a coupled one-dimensional climate, photochemistry, and spectral simulation model developed for terrestrial exoplanets, with the capability of inputting wavelength-dependent albedo profiles (following \citealt{Lin_2021MNRAS.505.3562L}; model described in detail in \citealt{Madden_2020_surfaces}). We simulate the high-resolution spectra with a line-by-line radiative transfer module, which was initially developed for studying the Earth’s atmosphere \citep{Traub_1976ApOpt..15..364T} and later modified to simulate the emission and transmission spectra of rocky exoplanets \citep[e.g.,][]{kaltenegger_spectral_2007, kaltenegger_transits_2009}. Our model includes the opacities of the major spectroscopically active molecules: C$_2$H$_6$, CH$_4$, CO, CO$_2$, H$_2$CO, H$_2$O, H$_2$O$_2$, H$_2$S, HNO$_3$, HO$_2$, N$_2$O, N$_2$O$_5$, NO$_2$, O$_2$, O$_3$, OCS, OH, and SO$_2$ from the HITRAN 2016 line lists \citep{Gordon2017}. Rayleigh scattering is also included and scaled to cool stars following \cite{Paradise_2021Icar..35814301P}.

We model both a modern Earth-analog atmosphere and a prebiotic Earth-analog atmosphere for TRAPPIST-1e. The modern Earth atmosphere has 1 bar surface pressure and modern Earth-like outgassing rates \citep[see e.g.,][]{Rugheimer_2013AsBio..13..251R}. The prebiotic atmosphere mimics the ancient Earth atmosphere about 3.9 billion years ago before the Great Oxidation Event \citep[see e.g.,][]{kaltenegger_high-resolution_2020}, which had 10 per cent CO$_2$ and only trace amount of O$_2$ and O$_3$. For the modern Earth scenario, we further model an atmosphere with 0.5 bar surface pressure to account for atmospheric erosion due to stellar activity \citep[e.g.,][]{airapetian_how_2017, bolmont_water_2017, bourrier_temporal_2017, dong_atmospheric_2018}. Both modern Earth TRAPPIST-1e atmosphere models assume 100 times higher CO$_2$ mixing ratio compared to PAL to keep the surface temperature above freezing. Key model parameters are summarized in Table \ref{tab:surface_parameters}. Temperature profiles and chemical mixing ratio profiles for CH$_4$, H$_2$O, O$_3$, and N$_2$O are presented in Figure \ref{fig:temp_mr_profile}. Note that while our models predict low O$_3$ concentration and no thermal inversion layer, which agrees with previous studies assuming red stars \citep[e.g.,][]{segura_ozone_2003, Rugheimer_2013AsBio..13..251R, rugheimer_spectra_2018}, \cite{jaziri_adam_yassin_developing_2021} predicted a concentrated O$_3$ layer that produces thermal inversion. The key difference is the assumed SED. \cite{jaziri_adam_yassin_developing_2021} adopted Galaxy Evolution Explorer (GALEX, \citealt{galex_2005ApJ...619L...1M}) far-UV and near-UV photometry (following \citealt{peacock_how_2018}), where the TRAPPIST-1 flux near 200 nm is comparable to modern Sun. However, here we follow the semiempirical SED from \cite{Wilson_2021ApJ...911...18W}, which predicts that TRAPPIST-1 emits $\sim1000$ times less flux than modern Sun at the same wavelength. Atmospheric O$_3$ is produced and destroyed via the Chapman reactions \citep{chapman_ozone_1930}, which are sensitive to high-energy radiation shortward of 240 nm and 320 nm, respectively. Therefore, SED assumptions, especially in the short ($\lesssim 200$ nm) wavelengths, would have profound impact on the O$_3$ profile and hence the temperature structure. For more details about the models, refer to \cite{Lin_2021MNRAS.505.3562L}.

\begin{figure*}
	\includegraphics[width=0.9\textwidth]{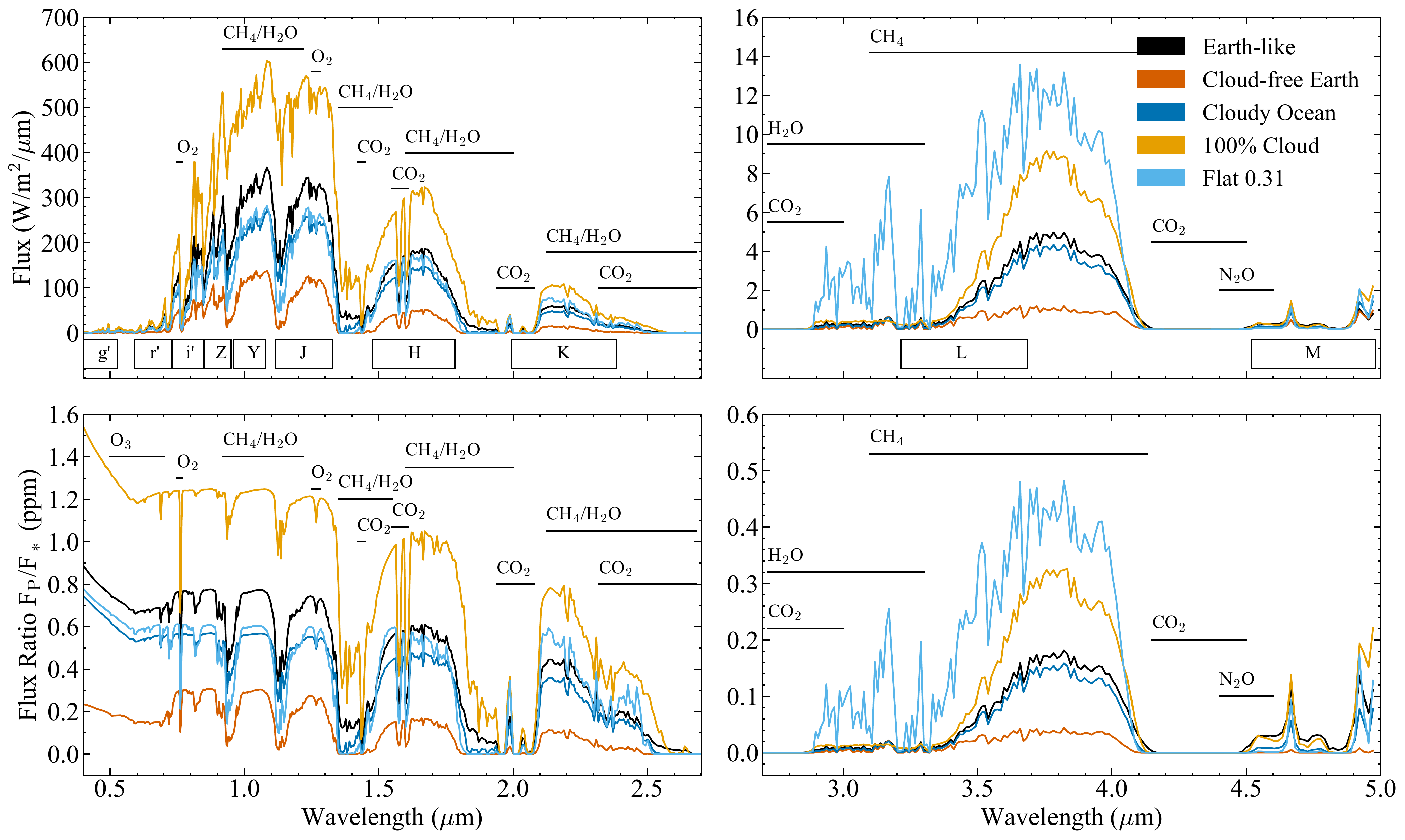}
    \caption{Reflection spectra for 1 bar modern Earth TRAPPIST-1e models with four surface compositions shown at R $= 300$. The spectra are shown in two units: (top row) absolute flux and (bottom row) planet-to-star flux contrast ratio ($F_{\rm p}/F_*$). Modern Earth-like surface with 44 per cent cloud coverage is shown in black, cloud-free modern Earth-like surface in red, ocean planet with 44 per cent cloud coverage in blue, 100 per cent cloudy planet in yellow, and a planet with flat 0.31 albedo in sky blue. The spectra are shown in two wavelength ranges: 0.4--2.7 $\micron$ (left) and 2.7--5.0 $\micron$ (right). The two wavelength ranges are shown in different scales for clarity and major spectral features are labeled. Ranges of common photometric bands are shown below the spectra. Presence of clouds significantly increases the overall spectral flux. The flat 0.31 albedo is an acceptable approximation in visible, but overestimates the reflected flux in NIR.}
    \label{fig:reflection_1bar_modern}
\end{figure*}

Surface albedos can have strong impacts on the climates of rocky exoplanets and thus on their spectra \citep[e.g.,][]{lin_high-resolution_2019, Madden_2020_surfaces, Rushby_2020_effect_of_land_albedo}. To demonstrate albedo effects on the planet's climate and spectrum, and to showcase the necessity to include wavelength-dependent albedo in modeling, we simulate four albedo scenarios for each atmospheric scenario: 
\begin{enumerate}
    \item Cloudy Earth-like planet, which is covered with 70 per cent liquid water ocean and 30 per cent land. The land surface is further divided into 30 per cent grass, 30 per cent trees, 9 per cent granite, 9 per cent basalt, 15 per cent snow, and 7 per cent sand. We assume 44 per cent of the planet is covered by a homogeneous opaque cloud layer located at 6 km, corresponding to the middle layer of Earth’s cloud layers \citep[following][]{kaltenegger_spectral_2007, Madden_2020_surfaces}.
    \item  Cloud-free Earth-like planet with aforementioned surface composition.\footnote{Note that cloud-free prebiotic model becomes too hot that it does not converge.}
    \item Cloudy ocean planet, where the surface is completely covered by water ocean, and a homogeneous opaque cloud layer at 6 km covers 44 per cent of the planet.
    \item Fully cloudy planet, where the entire planet is covered by a homogeneous opaque cloud layer at 6 km and the surface is not accessible to remote observation.
    \item A planet with wavelength-independent, flat surface albedo of 0.31 and zero cloud coverage.
\end{enumerate}

Clouds affect the spectra and climate feedback of a planet by hiding the atmosphere and surfaces below and by changing the overall planetary albedo. The climate effects of clouds on tidally locked M dwarf planets have been studied by several teams with 3D general circulation model (GCM) simulations \citep[e.g.,][]{zhang_surface_2017, fauchez_impact_2019, fauchez_trappist-1_2020, sergeev_atmospheric_2020}, producing a range of interesting results that will deepen our understanding of energy effects of clouds and inform future observations on M-star planets like TRAPPIST-1e. While the 3D modeling efforts are ongoing, we explore the effect of water clouds similar to Earth’s on the spectra of TRAPPIST-1e models using three scenarios: cloud-free, partial clouds (44 per cent), and full cloud cover (100 per cent). Note that we treat clouds as a homogeneous opaque layer and set them at 6 km altitude, corresponding to the middle layer of Earth’s cloud layers \citep[following][]{kaltenegger_spectral_2007, Madden_2020_surfaces}.

We simulate the high-resolution reflection and emission spectra at a step size of 0.01 cm$^{-1}$ \citep[following][]{kaltenegger_transits_2009}. The reflection spectra cover the visible to near-infrared (NIR) wavelength range from 0.4 to 5 $\micron$. The emission spectra cover a wavelength range from 4 to 20 $\micron$. In the overlapping wavelength range (4--5 $\micron$), we sum the reflection and emission spectra. In visible and NIR, spectral resolution R $= \lambda / \Delta \lambda > 100,000$, exceeding the proposed resolution of ArmazoNes high Dispersion Echelle Spectrograph (ANDES) designed for the E-ELT. In IR, R $> 50,000$. Details of the radiative transfer model are discussed in \cite{Lin_2021MNRAS.505.3562L}.


\begin{figure*}
	\includegraphics[width=0.9\textwidth]{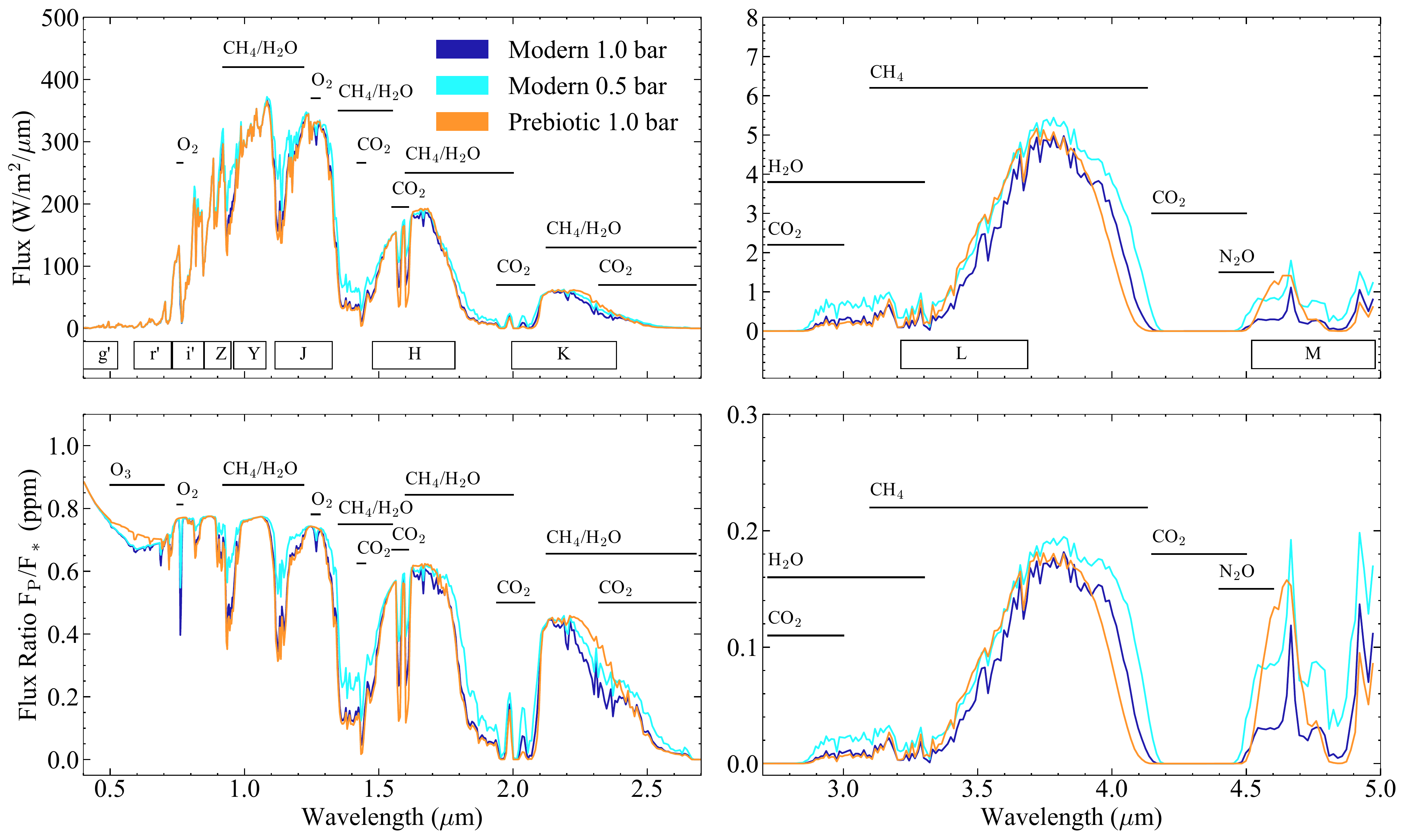}
    \caption{Reflection spectra for TRAPPIST-1e models with Earth-like surface composition shown at R $= 300$ in the same setup as Figure \ref{fig:reflection_1bar_modern}. The modern 1 bar model is shown in navy, the modern 0.5 bar model in cyan, and the prebiotic 1 bar model in orange. The 0.5 bar spectrum shows shallower features due to less absolute abundance of molecules. O$_2$ and O$_3$ features are absent from the prebiotic spectrum, but CO$_2$ features are stronger.}
    \label{fig:reflection_earth_surface}
\end{figure*}

\section{Results and Discussion} \label{sec:results_discussion}

To provide our spectra as tools for observers, they are shown in (i) absolute flux of the planet and (ii) planet to star flux ratio for reflection (Figure \ref{fig:reflection_1bar_modern} and \ref{fig:reflection_earth_surface}) and emission (Figure \ref{fig:emission_spectra_1barModern} and \ref{fig:emission_spectra_earthSurface}) spectra. The spectral models are calculated at high resolution (0.01 cm$^{-1}$ step size) and available online, but are smeared to R $= 300$ in the figures for clarity. 

To emphasize our main model results: the O$_2$ + CH$_4$ and O$_3$ + CH$_4$ biosignature pairs as well as climate indicators CO$_2$ and H$_2$O show features in both reflection and emission spectra of TRAPPIST-1e, using the updated atmosphere model with wavelength-dependent surface albedo and the recent Mega-MUSCLES stellar SED. The inclusion of wavelength-dependent albedo has significant impact on the overall flux of both reflection and emission spectra. Detectability of some spectral signatures is affected by surface reflectivity and presence of clouds.

\subsection{Modern Earth TRAPPIST-1e Spectra} \label{sec:modern_earth}

\subsubsection{Reflection Spectra}
In the modern Earth-like atmospheric scenarios (both 1 bar and 0.5 bar surface pressure), the reflection spectral models of TRAPPIST-1e are dominated by CO$_2$, H$_2$O, and CH$_4$ features. Surface composition and cloud distribution change the overall flux and depths of features, but not the overall atmospheric chemistry (Figure \ref{fig:temp_mr_profile}).

CO$_2$ shows features at 1.6, 2.0, 2.7, and 4.2 $\micron$. CO$_2$ also shows a feature at 1.4 $\micron$, but it is obscured by strong CH$_4$ and H$_2$O features. The 1.6 $\micron$ CO$_2$ feature is the most detectable among all, with a feature depth of roughly 0.6 ppm for the 1 bar fully cloudy model, 0.4 ppm for the partially cloudy models, and 0.1 ppm for the cloud-free model (Figure \ref{fig:reflection_1bar_modern}). Depth of this feature halves for the 0.5 bar models (Figure \ref{fig:reflection_earth_surface})

H$_2$O shows features at 1.1, 1.4, 1.9, and 2.7 $\micron$. Note that in the visible range, CH$_4$ and H$_2$O features overlap, making it challenging to distinguish the two molecules at low resolution \citep[see also][]{Kaltenegger_Finding_2020ApJ...904...10K}. 
The most prominent features of CH$_4$ are at 1.4, 1.9, 2.3, and 3.3 $\micron$. Two weaker CH$_4$ features are present at about 0.9 and 1.1 $\micron$.
CH$_4$-H$_2$O combined features around 1.4 and 1.9 $\micron$ provide the best opportunity to detect these two species. These two features are approximately 1 ppm deep for the fully cloudy model, 0.6 ppm for the 1 bar partially cloudy models, and 0.2 ppm for the cloud-free model. The partially cloudy and cloud-free models show weaker features in part due to saturation (Figure \ref{fig:reflection_1bar_modern}). 0.5 bar model feature depths are roughly 80 per cent of the 1 bar model feature depths (Figure \ref{fig:reflection_earth_surface}).

In the optical, O$_2$ has a feature at 0.76 $\micron$ and O$_3$ shows a broad continuum at 0.45--0.74 $\micron$ (Chappuis band), which can be seen when plotting the planet-to-star contrast ratio (Figure \ref{fig:reflection_1bar_modern} and Figure \ref{fig:reflection_earth_surface}, bottom row). The 0.76 $\micron$ O$_2$ feature is 0.6, 0.4, and 0.2 ppm deep for the 1 bar fully cloudy, partially cloudy, and cloud-free albedo scenarios, respectively (Figure \ref{fig:reflection_1bar_modern}), while the feature depth in the 0.5 bar spectrum is roughly halved (Figure \ref{fig:reflection_earth_surface}). The O$_3$ Chappuis band is no stronger than 0.1 ppm for all models, and partially modified by the Rayleigh scattering slope. In absolute flux, the absorption features of O$_2$ and O$_3$ in reflected light are strongly modified by the stellar absorption lines, making them challenging to identify (Figure \ref{fig:reflection_1bar_modern} and Figure \ref{fig:reflection_earth_surface}, top row).

N$_2$O has also been proposed as a biosignature \citep[e.g.,][]{Lederberg_1965Natur.207....9L, Segura_2005AsBio...5..706S, Rugheimer_2013AsBio..13..251R, Grenfell_2014P&SS...98...66G}, even though some argues that stellar flares lead to efficient abiotic N$_2$O production \citep[][]{airapetian_how_2017}. In our modern Earth scenarios, N$_2$O has a $\sim 4 \times 10^{-7}$ surface mixing ratio. The strongest N$_2$O feature is located at 4.5 $\micron$, but is obscured by an adjacent strong CO$_2$ feature. Weaker N$_2$O features are located around 2.9, 3.6, and 3.9 $\micron$, but all are obscured by stronger features nearby.

\subsubsection{Emission Spectra}

\begin{figure*}
	\includegraphics[width=0.9\textwidth]{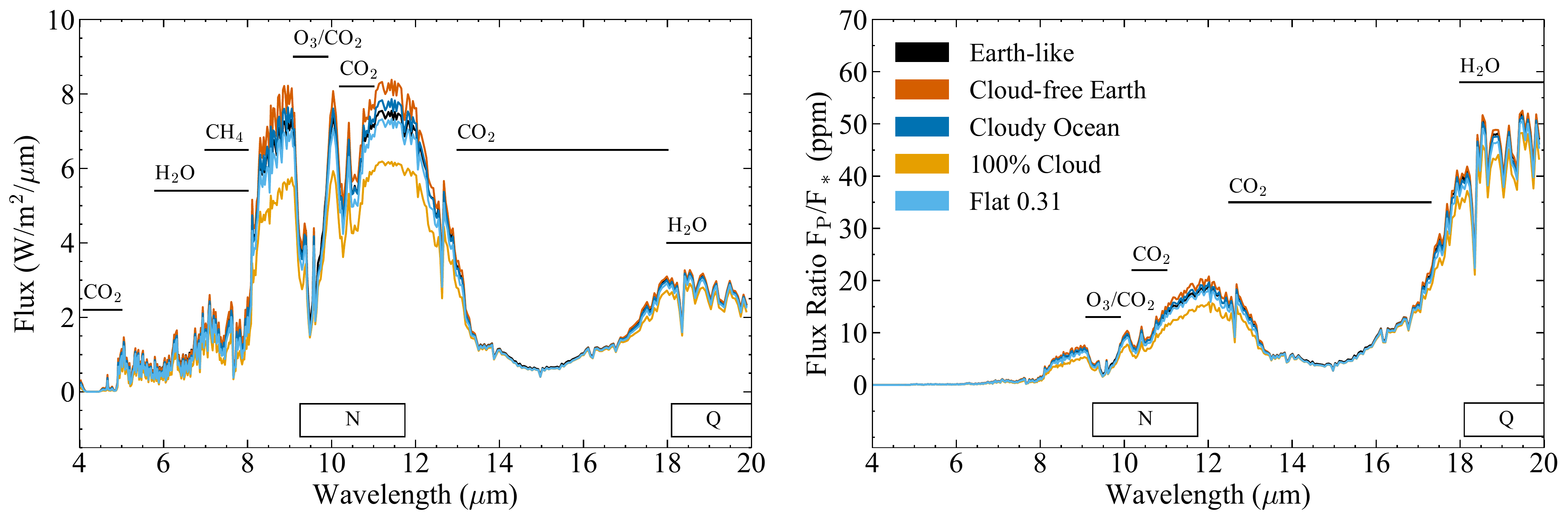}
    \caption{Emission spectra for 1 bar modern Earth TRAPPIST-1e models with four surface compositions shown at R $= 300$. The spectra are shown in two units: (left) absolute flux and (right) planet-to-star flux contrast ratio ($F_{\rm p}/F_*$). Modern Earth-like surface with 44 per cent cloud coverage is shown in black, cloud-free modern Earth-like surface in red, ocean planet with 44 per cent cloud coverage in blue, 100 per cent cloudy planet in yellow, and a planet with flat 0.31 albedo in sky blue. Major spectral features are labeled. Ranges of common photometric bands are shown below the spectra. Albedo affects emission spectra indirectly by changing the temperature and therefore the blackbody profile of the emitting layer. The warmest cloud-free model has the highest flux, while the most reflective, coolest 100 per cent cloud model has the lowest.}
\label{fig:emission_spectra_1barModern}
\end{figure*}

\begin{figure*}
	\includegraphics[width=0.9\textwidth]{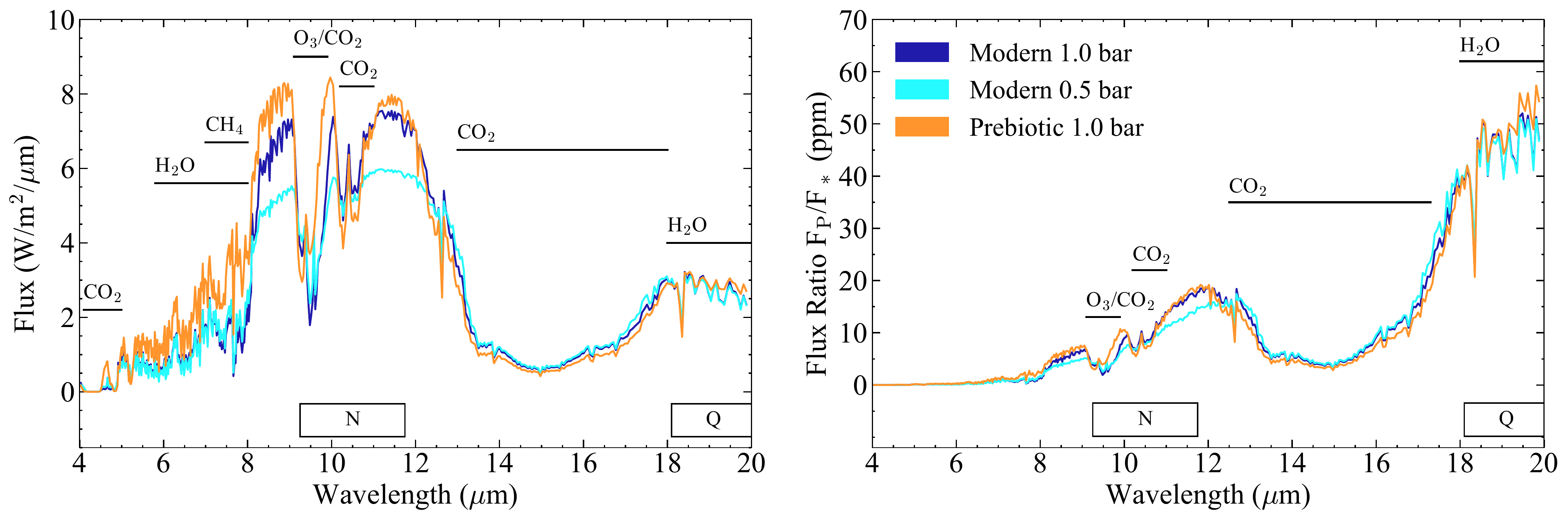}
    \caption{Emission spectra for TRAPPIST-1e models with Earth-like surface composition shown at R $= 300$ in the same setup as Figure \ref{fig:emission_spectra_1barModern}. The modern 1 bar model is shown in navy, the modern 0.5 bar model in cyan, and the prebiotic 1 bar model in orange. Excessive greenhouse effect from 10 per cent CO$_2$ makes the prebiotic Earth model the warmest, so it emits the highest flux.}
\label{fig:emission_spectra_earthSurface}
\end{figure*}

Figure \ref{fig:emission_spectra_1barModern} and Figure \ref{fig:emission_spectra_earthSurface} show the emission spectra and the most prominent IR features for models of TRAPPIST-1e as both absolute flux and planet-to-star contrast ratio from 4 to 20 $\micron$. The modern Earth-like atmospheric scenario is dominated by CO$_2$, H$_2$O, and O$_3$ emission features. 

CO$_2$ shows a broad feature centered at 15 $\micron$ with wings extending to about 12 and 18 $\micron$ on both sides, and two other features at 4.2 and 10.3 $\micron$. CO$_2$ also has a feature around 9.4 $\micron$ that is partially obscured in low resolution by the stronger 9.6 $\micron$ O$_3$ feature in both modern models, but can be distinguished in high-resolution.
The 12--18 $\micron$ band has a depth of roughly 15--40 ppm, while the 10.3 $\micron$ feature has a depth of roughly 4 ppm. The 4.2 $\micron$ feature is not detectable due to low flux in short wavelength.

H$_2$O shows a feature a 6.4 $\micron$ and a continuum at $> 18\,\micron$. 
The 6.4 $\micron$ feature is not detectable due to low flux in short wavelength, but the continuum has approximately 10--20 ppm variations and is potentially detectable.
A CH$_4$ feature is present at 7.5 $\micron$, but partially overlaps with a nearby water feature and is too shallow to be detectable in low resolution.

O$_3$ shows a strong saturated feature at 9.6 $\micron$ ($\sim6$ ppm deep) and a weaker feature at 4.7 $\micron$ (not detectable in low resolution). Note that the prebiotic Earth model has negligible amount of O$_3$ but shows a CO$_2$ feature at this wavelength.
The strongest N$_2$O features are located around 4.5, 7.9 and 17 $\micron$, but are challenging to detect because the 7.9 $\micron$ feature overlaps with CH$_4$ and H$_2$O features and the 17 $\micron$ feature overlaps with CO$_2$ and H$_2$O features.

The strength of all spectral absorption features decrease with decreasing chemical abundance for the eroded atmosphere model with a surface pressure of 0.5 bar. While the mixing ratios are similar, the absolute abundance decreases due to the overall surface pressure decrease from 1 bar to 0.5 bar (see \citealt{lin_high-resolution_2019}). In IR, the cooler surface temperature of the 0.5 bar model results in lower emission flux of the planet model (see Table \ref{tab:surface_parameters}).

\subsubsection{Spectral Biosignatures}

Simultaneous detection of O$_2$ + CH$_4$ or O$_3$ + CH$_4$ are considered strong biosignatures \citep[e.g.,][]{Lederberg_1965Natur.207....9L, Lovelock_1965Natur.207..568L, Lippincott_1967ApJ...147..753L, Sagan_1993Natur.365..715S}. Assuming modern Earth-like biogenic emission rates, these two biosignature pairs show features in both the reflection and the emission spectra of TRAPPIST-1e.

The CH$_4$-H$_2$O compound features at 1.4 and 1.9 $\micron$, if observed in combination with the 0.76 $\micron$ O$_2$ feature, constitute the strongest biosignature pairs in the visible to NIR range (0.4--5 $\micron$). High-resolution spectroscopy is likely required to differentiate CH$_4$ from H$_2$O. The 9.6 $\micron$ O$_3$ feature in combination with the 7.5 $\micron$ CH$_4$ feature is the strongest spectral biosignature pair in the infrared range (5--20 $\micron$) for our modern Earth-like atmospheric composition TRAPPIST-1e models, but the 7.5 $\micron$ CH$_4$ feature has low flux ratio and may require high-resolution observation to detect. Collision-induced absorption (CIA) contributes to the absorption spectra and provides an alternative venue to constrain atmospheric O$_2$ \citep{karman_update_2019}. \textsc{Exo-Prime2} does not include CIA data. For CIA effects on TRAPPIST-1e, refer to \cite{Lin_2021MNRAS.505.3562L} where transmission spectra with CIA opacities were simulated using the \textsc{POSEIDON} atmospheric retrieval code \citep{macdonald_hd_2017}.

\subsection{Prebiotic Earth TRAPPIST-1e Spectra} \label{sec:prebiotic_earth}
In the prebiotic Earth TRAPPIST-1e scenario, the spectra show several large differences due to the difference in atmospheric composition and temperature (Figure \ref{fig:reflection_earth_surface} and \ref{fig:emission_spectra_earthSurface}), providing a key to differentiate between the modern and prebiotic Earth atmospheric scenarios. O$_2$ and O$_3$ features are absent in the prebiotic model with anoxic atmosphere (Figure \ref{fig:temp_mr_profile}). The prebiotic model also shows slightly weaker CH$_4$ absorption features compared to the modern Earth TRAPPIST-1e model due to lower CH$_4$ mixing ratio (Figure \ref{fig:temp_mr_profile}). CO$_2$ absorption features are deeper (by at most $\sim0.1$ ppm in reflection and $\sim3$ ppm in emission) due to higher CO$_2$ mixing ratio. Spectral absorption features of H$_2$O are comparable between the modern Earth TRAPPIST-1e model and the prebiotic scenario, for surface pressures of 1 bar. In IR, the prebiotic Earth TRAPPIST-1e spectral model shows higher flux due to higher surface temperature. 

The authors showed that prebiotic and modern Earth TRAPPIST-1e atmospheric scenarios can also be distinguished with \textit{JWST} transmission observations \citep{Lin_2021MNRAS.505.3562L}. ELT reflected light observations can potentially differentiate between these scenarios too,  but further investigation using instrument characteristics is needed, once those become available.


\subsection{Effects of Wavelength-dependent Albedo}
We model five albedo scenarios -- cloudy Earth-like, cloud-free Earth-like, cloudy ocean world, fully cloudy planet, and flat 0.31 albedo -- and overplot their reflection (Figure \ref{fig:reflection_1bar_modern}) and emission (Figure \ref{fig:emission_spectra_1barModern}) spectra for the 1 bar modern Earth atmospheric composition. The impact of wavelength-dependent albedo on reflection and emission spectra is clearly visible.

In reflected light, albedo has significant impact on the overall flux of the spectra (Figure \ref{fig:reflection_1bar_modern}). Both absolute flux and planet-to-star contrast ratio are proportional to planetary albedo. The fully cloudy model nearly doubles the flux of the partially cloudy models, and roughly quadruples the flux of the cloud-free model. The flat 0.31 albedo is an acceptable approximation in the visible, where it is broadly consistent with the cloudy Earth-like surface and cloudy ocean planet models. In NIR, however, it significantly overestimates reflected flux because realistic planetary surfaces are less reflective in longer wavelengths.

Clouds have both negative and positive impact on the detectability of spectral signatures \citep[see also][]{kaltenegger_spectral_2007}. The CH$_4$ and H$_2$O combined features at 0.9 and 1.1 $\micron$ are slightly weaker on the fully cloudy model spectrum, because water is concentrated near the surface where evaporation takes place (Figure \ref{fig:temp_mr_profile}), and clouds obstruct the atmosphere below 6 km. The O$_3$ band near 0.6 $\micron$ and the O$_2$ feature at 0.76 $\micron$, however, are deeper on the 100 per cent cloudy spectrum. This is because O$_3$ mixing ratio increases towards upper atmosphere while O$_2$ is well-mixed throughout the atmosphere with a 0.21 mixing ratio. Absorption above 6 km plays a more important role for these oxygenic species, so the effect of increased reflected flux counteracts the negative effect of clouds. Finally, the CH$_4$-H$_2$O-CO$_2$ combined features around 1.4 and 1.9 $\micron$ also show an increased relative depth due to the increased albedo of a cloudy planet.

Emitted flux does not scale with albedo directly, but surface composition and cloud configuration can shape emission spectra indirectly by influencing the atmospheric energy budget. The least reflective, cloud-free model efficiently absorbs stellar irradiation and has a surface temperature of 296 K, while the 100 per cent cloudy model is 18 K cooler at surface (Table \ref{tab:surface_parameters}). Emitted flux, which scales with the blackbody radiation of the emitting layer, is greatest for the cloud-free model (Figure \ref{fig:emission_spectra_1barModern}).

\subsection{Cloud Distribution on Tidally-locked TRAPPIST-1e}
By assuming two extreme cases (cloud-free and 100 per cent cloudy) and a partially cloudy case, we explore the effects of clouds on TRAPPIST-1e reflection and emission spectra using our 1D \textsc{Exo-Prime2} model. Close-in planets such as TRAPPIST-1e are likely tidally locked \citep{turbet_modeling_2018}. On synchronously rotating planets, 3D dynamic spatial distribution of clouds can have significant impact on energy balance that cannot be fully captured by 1D models. Using 3D GCM, \cite{yang_stabilizing_2013} showed that thick clouds would form near the substellar point on tidally locked planets, shielding such planets from excessive irradiation and extends the inner edge of HZ around red stars. For planets with short rotation periods such as TRAPPIST-1e ($\approx6.1$ days, \citealt{agol_refining_2021}), \cite{kopparapu_inner_2016} predicted that rapid rotation would shift the substellar cloud deck eastward and smear clouds around the planet to form banded structure. Simulation by \cite{fauchez_impact_2019} produced a thick cloud deck extending to the terminator, agreeing with previous studies.

Note that 3D GCM incorporating albedo feedback showed that TRAPPIST-1e is the most habitable planet in the TRAPPIST-1 system. Surface liquid water ocean can exist (at least around the substellar point) under a wide array of atmospheric scenarios \citep{wolf_assessing_2017, turbet_modeling_2018, fauchez_impact_2019}. Therefore, our albedo choices of Earth-like surface and ocean world surface are physically justified.

\subsection{Implications for ELT and \textit{JWST}} \label{sec:observe_implications}

Here, we present the high-resolution reflection and emission spectral models for one of the most favorable targets for the search for biosignature gases on rocky exoplanets -- TRAPPIST-1e. We model both modern Earth TRAPPIST-1e atmospheres with 1 bar surface pressure and a 0.5 bar model to account for atmospheric erosion by M dwarfs. We also model a prebiotic Earth-like scenario to account for the evolutionary history of a terrestrial planet. We improve the models presented in \cite{lin_high-resolution_2019} using the recently available Mega-MUSCLES SED for TRAPPIST-1 as stellar input and also model the planetary atmosphere with the updated \textsc{Exo-Prime2} model, which incorporates wavelength-dependent surface albedo. To demonstrate the effects of wavelength-dependent albedo, for each atmospheric composition, we model five albedo scenarios -- cloudy Earth-like, cloud-free Earth-like, cloudy ocean world, fully cloudy planet, and flat 0.31 albedo.

Our spectral models show that CH$_4$, CO$_2$, and H$_2$O features are present in both reflection and emission spectra of TRAPPIST-1e. O$_2$ + CH$_4$ and O$_3$ + CH$_4$ are the most prominent spectral biosignatures present in reflection and emission spectral models, respectively. Quantitative constraints on their detectability requires future modeling that accounts for instrument characteristics. Our simulated spectra also show that surface albedo has a strong impact on the overall reflected and emitted flux from TRAPPIST-1e, and their detectability. Absorption features in the visible are more affected than features in the IR, as Figure \ref{fig:reflection_1bar_modern} and Figure \ref{fig:emission_spectra_1barModern} show for the O$_3$, O$_2$, and CH$_4$ features, which are constituents of spectral biosignature pairs. In the visible the absorption features show larger changes in depth due to different surface reflectivity.

The angular separation of TRAPPIST-1e from its host star is 2.4 milliarcsecond (mas). ELT is expected to achieve a 6 mas inner working angle, thus TRAPPIST-1e won't be resolved by ELT. However, detecting spectral features on unresolved planets is possible \citep[e.g.,][]{Brogi_2014A&A...565A.124B}.
Such observation can complement spectral observations of TRAPPIST-1e in transit with \textit{JWST}, where abundances for key molecules including CO$_2$, CH$_4$ and H$_2$O can be constrained within a 20-transit campaign assuming modern Earth-like atmosphere \citep[e.g.,][]{Lin_2021MNRAS.505.3562L}. Measuring the thermal emission spectra of TRAPPIST-1e via secondary eclipse observations, however, will be much more challenging. Due to its cool temperature, over 100 visits are required to detect the secondary eclipse of TRAPPIST-1e, making such observation impractical over the lifetime of \textit{JWST} \citep{morley_observing_2017}.

Our high-resolution database for modern and prebiotic Earth TRAPPIST-1e models provides a tool for observers to train retrieval algorithms and plan observation strategies to characterize TRAPPIST-1e. 
The combination of transmission spectroscopy with \textit{JWST} and reflection plus emission spectroscopy with ELT can probe the atmosphere as well as the surface, which will provide a powerful tool to perform atmospheric reconnaissance for TRAPPIST-1e with the potential to uncover the first signs of life on another world.

\section*{Acknowledgements}
Z.L. acknowledges support from the MIT Presidential Fellowship. L.K. acknowledges funding from the Brinson Foundation and the Carl Sagan Institute at Cornell University.
\section*{Data Availability}

All high-resolution spectral models are available online at \href{https://doi.org/10.5281/zenodo.6365743}{https://doi.org/10.5281/zenodo.6365743}.
 



\bibliographystyle{mnras}
\bibliography{main} 




\appendix




\bsp	
\label{lastpage}
\end{document}